\documentclass[aps,twocolumn,eqsecnum]{revtex4}
\usepackage{graphics}

\begin{document}


\title{The $\theta^+$ baryon in soliton models: large $N_c$ QCD
and the validity of  rigid-rotor quantization}

\author{Thomas D. Cohen} \email{cohen@physics.umd.edu}

\affiliation{Department of Physics, University of Maryland,
College Park, MD 20742-4111}

\begin{abstract}
A light collective $\theta^+$ baryon state (with strangeness +1)
was predicted via rigid-rotor collective quantization of SU(3)
chiral soliton models.  This paper explores the validity of this
treatment.  A number of rather general analyses suggest that
predictions of exotic baryon properties based on  this
approximation do not follow from large $N_c$ QCD. These include
an analysis of the baryon's width, a comparison of the
predictions with general large $N_c$ consistency conditions of
the Gervais-Sakita-Dashen-Manohar type; an application of the
technique to QCD in the limit where the quarks are heavy; a
comparison of this method with the vibration approach of Callan
and Klebanov; and the $1/N_c$ scaling of the excitation energy.
It is suggested that the origin of the problem lies in an
implicit assumption in the that the collective motion is
orthogonal to vibrational motion. While true for non-exotic
motion, the Wess-Zumino term induces mixing at leading order
between collective and vibrational motion with exotic quantum
numbers. This suggests that successful phenomenological
predictions of $\theta^+$ properties based on rigid-rotor
quantization were accidental.
\end{abstract}


\maketitle

\section{introduction\label{Intro}}
A narrow baryon resonance with a strangeness of +1 ({\it i.e.}
containing one excess strange antiquark) has recently been
identified by a number of experimental groups\cite{exp}. Such a
state is exciting: it is unambiguously exotic in the sense that it
cannot be a simple three-quark state. These experiments have
spurred considerable theoretical activity. Much of this activity
has been aimed at understanding the structure of this exotic
state.  The most common treatment of this problem has been based
on variants of the quark model; the new baryon is identified as a
pentaquark\cite{qm}. Other approaches treat the $\theta^+$ in
terms of  meson-baryon binding\cite{IKOR,KK} or as a kaon-pion
nucleon state\cite{LOM} or are based on on QCD sum-rules
\cite{sumrules}.  While these approaches are all interesting,
they are are also all highly model dependent and it is difficult
to assess in an {\it a priori} way their validity. A discussion
of some the issues raised by various models may be found in
\cite{JM}.

The analysis based on the SU(3) chiral soliton model treated with
rigid-rotor quantization
\cite{Pres,DiaPetPoly,WalKop,BorFabKob,Kim} appears different
from other treatments of the $\theta^+$ structure in a number of
ways : i) The approach was used to predict the state and its
properties \cite{Pres,DiaPetPoly}.  In contrast most of the other
treatments were subsequent to the experimental discoveries. ii)
The predictions of the mass were very
accurate\cite{Pres,DiaPetPoly}; the width was predicted to be
narrow\cite{DiaPetPoly}, which is consistent with the widths
presently observed \cite{Nus}.  iii) The prediction of the mass
is totally insensitive to the dynamical details of the model. The
analysis holds regardless of the detailed shape of the profile
function that emerges from the dynamics.

While the first two points are certainly of interest, the third
point is of far greater significance from the point of view of
theory. It was noticed quite some time ago by Adkins and Nappi
that a number of relationships between observables in chiral
soliton models depend on the structure of the models and their
symmetries, but do not depend on the dynamical details such as the
parameters in the lagrangian \cite{AdkNap}. Moreover, these
relations were derived in a fully model-independent way directly
from large $N_c$ using consistency relations\cite{GS,DM,Jenk,DJM}.

In light of this, it is natural to conjecture that the
relationships between masses underlying the analysis in
refs.~\cite{Pres,DiaPetPoly,WalKop,BorFabKob,Kim} which do not
depend on the dynamical details of the model are similarly
model-independent consequences of large $N_c$ QCD.  If this is
true, it greatly aids in our understanding of the $\theta^+$:
modulo corrections due to $1/N_c$ effects of higher order effects
in SU(3), the $\theta^+$ seems essentially understood. However, it
remains quite controversial as to whether this is indeed the case.

In a previous paper\cite{Coh03}, it was argued that the prediction
of properties of exotic states based on rigid-rotor quantization
are {\it not} generic predictions of large $N_c$ QCD. Before
discussing the content of this paper, a linguistic issue should
be addressed to avoid possible confusion. The analysis is done in
a $1/N_c$ expansion so one must be able to consider systems with
varying $N_c$. Accordingly throughout this paper the phrases
``exotic state'' or ``exotic baryon'' refer to states which are
exotic for the given value of $N_c$ which is relevant (and not
necessarily for $N_c=3$); {\it i.e.,} those baryons which cannot
be described in a quark model with $N_c$ quarks. Clearly states
with positive strangeness are exotic for any $N_c$.

Two basic arguments were presented in ref.~\cite{Coh03} which
suggest that the rigid-rotor quantization is not valid for exotic
states. The first was based on large $N_c$ consistency rules of
the sort discussed above. {\it All} of the previously known model
independent properties of baryons were derived from such
relations. However, such relations do not predict the existence
of collective exotic states at large $N_c$. The only collective
states predicted are precisely those with the same quantum
numbers as those which emerge in a large $N_c$ quark model with
all quarks in the lowest s-wave orbital. The second argument was
that the excitation energies of the exotic state of order $N_c^0$
is inconsistent with the type of scale separation needed to
justify collective quantization.

Itzhaki,  Klebanov, Ouyang and  Rastelli (IKOR)\cite{IKOR} reached
similar conclusions about the validity of the rigid rotor
approximation for exotic states from a rather different starting
point: the treatment of SU(3) symmetry breaking. It is clear that
for large symmetry breaking fluctuations into the strangeness
direction are of a vibrational character and the appropriate
formalism to describe these are the Callan-Klebanov ``bound
state'' approach\cite{CK}. For the case of non-exotic baryons it
can be shown that as the SU(3) symmetry breaking goes to zero, the
predicted energies and spatial distributions in the
Callan-Klebanov approach match those of the rigid-rotor
quantization\cite{CK}; the two approaches are compatible for
non-exotic states in the regime where both should work.  In
contrast, at small but nonzero SU(3) symmetry breaking the
excitation energy for exotic states such as the $\theta^+$ as
calculated via the rigid-rotor treatment does not match smoothly
onto the Callan-Klebanov value.  IKOR took this as evidence that
rigid-rotor treatment, while valid for non-exotic states, gives
spurious results for exotic states.

Pobylitsa\cite{Pob} also concluded that rigid-rotor quantization
failed for exotic states based on a study of an exactly solvable
toy model of the Lipkin-Meshkov-Glick type \cite{LMG}.

Despite the arguments in refs.~\cite{Coh03,IKOR,Pob}, it is not
universally accepted that rigid-rotor collective quantization of
chiral soliton models is invalid for exotic baryon properties.
Diakonov and Petrov\cite{DP} have recently argued that that
rigid-rotor collective quantization  is accurate for such states
up to $1/N_c$ corrections, and have specifically attempted to
rebut the argument in ref.~\cite{Coh03}. Both a general analysis
as to why the authors believe the argument of ref.~\cite{Coh03}
is incorrect and a toy model to illustrate the point are given.
However, as will be discussed below, the general arguments are
flawed and the toy model is not anlagous to the problem of
interest.

It is hoped that the present paper will end help this controversy
by giving convincing evidence that the rigid-rotor collective
quantization is not valid at large $N_c$.  Given this goal, it is
important to state precisely what is meant by this claim so as to
minimize possible misunderstandings. The claim is formal. If
rigid-rotor quantization were valid at large $N_c$, then
predictions based on it would become arbitrarily accurate as $N_c
\rightarrow \infty$ ({\it i.e.,} there are no corrections due to
other effects which survive at large $N_c$).  Arguments are given
here that this is false: even at large $N_c$ there are nonzero
corrections. This is in contrast to properties of non-exotic
baryons for which calculations based on rigid-rotor quantization
do become arbitrarily accurate.

It is important to make quite clear the fundamental nature of the
difficulty with the rigid-rotor collective quantization. The issue
is not related to whether this treatment based on large $N_c$
analysis is a good approximation to the physical world of $N_c=3$,
but to the nature of exotic states in the large $N_c$ world.

This paper presents evidence for the invalidity of the
rigid-rotor quantization for exotic baryons.  Several general
arguments are presented showing that the rigid-rotor quantization
leads to predictions of exotic baryon properties which are
inconsistent with known large $N_c$ results.  These general
arguments strongly suggest an inconsistency between various
predictions of exotic baryon properties at large $N_c$ as based
on rigid-rotor quantization and the known behavior at large
$N_c$.  If correct, they imply  a flaw in the original derivation
of the rigid-rotor collective quantization of
refs.~\cite{SU3Quant} for these exotic states. Such a flaw must
give spurious results for exotic states while being valid for
non-exotic states. An analysis of the derivation indicates where
such a flaw may lie. The derivation implicitly assumes the
orthogonality of collective and vibrational motion. However, the
Wess-Zumino term induces a coupling between vibrational modes and
the collective motion associated with exotic excitations which
spoils this orthogonality.

This paper is organized as follows:  In the following section the
salient features of the treatment of exotic baryons based on the
rigid-rotor collective
quantization\cite{Pres,DiaPetPoly,WalKop,BorFabKob,Kim} is
briefly presented.  The next section discusses the large $N_c$
scaling behavior of the mass splittings between exotic states and
the ground state.  Following this, a series of arguments
arguments are presented that indicate that the collective
quantization treatment yields results in conflict to what one
expects from general large $N_c$ considerations.   A section
detailing a possible flaw in the original derivation of
rigid-rotor quantization as applied to exotic states is next,
followed by  a discussion of the claims of ref.~\cite{DP} which
purports to demonstrate the validity of rigid rotor.  The final
section discusses  the significance of these results in light of
the recent experimental reports of exotic s=1 baryons.

\section{Rigid-rotor predictions for exotic baryons \label{Rigid}}

There are a number of important assumptions which go into the
predictions of exotic baryon states in
refs.~\cite{Pres,DiaPetPoly,WalKop,BorFabKob,Kim}.  These include
the assumption that low-order perturbation theory in the strange
quark mass is justified for real world values, the assumption
that  $1/N_c$ expansion truncated at low order is justified for
these observables for $N_c=3$ as well as the assumption that
rigid-rotor quantization is valid for exotic states.  This paper
focuses on the issue of the validity of rigid-rotor quantization.
It is worth bearing in mind, however, that these other
assumptions are not totally innocuous. For example, prior to the
experimental observation of the $\theta^+$, Weigel\cite{Weigel}
observed that effects which were higher-order in the strange
quark mass induced mixing between the vibrational and collective
modes which had nontrivial effects on predictions of the
properties of the exotic states.  However, the central questions
of principle addressed in this paper are seen at leading order in
$1/N_c$ and in the exact SU(3) limit and we review the leading
order treatment below.

The analysis is based on a standard treatment of SU(3) chiral
soliton models developed in the mid-1980s\cite{SU3Quant}. The
starting point in the analysis is a classical static ``hedgehog''
configuration in an SU(2) subspace (which for convenience one may
take to be the u-d subspace). The profile function of this
hedgehog is obtained by minimizing the action for a static
configuration subject to the constraint that the baryon number
(which is taken to be the topological winding number for the
chiral field) is unity.  The detailed shape of the profile
function depends on the model---the types of couplings included in
the values of the parameters, and so on. However, the general
structure of the theory is completely model independent.  As
noted above, for the present purpose it is sufficient to consider
the exact SU(3) symmetric limit of the theory.  In the absence of
symmetry breaking effects there are eight collective (rotational)
variables.  That is, there are eight flat directions in which one
can rotate to the classical configuration to obtain a new
classical configuration which also corresponds to static
solutions of the classical equations of motion.

These collective variables are then quantized semi-classically
using an SU(3) generalization\cite{SU3Quant} of the usual SU(2)
collective quantization scheme\cite{ANW}.  If the classical
collective motion is fully decoupled from the internal motion of
the hedgehog shape, then one can quantize the two motions
separately.  Assuming this to be true, the dynamics of the
collective variables are expressed in terms of time-dependent
SU(3) rotations on the hedgehog shape
\begin{equation}
U(\vec{x},t) = R(t) U_0(\vec{x}) R^{\dagger}(t) \label{ut}
\end{equation}
where $R(t)$ is a global (space-independent) time-dependent SU(3)
rotation,  and $U_0$ is the hedgehog solution.  Inserting this
equation into the full lagrangian yields a collective lagrangian
whose variables are $R$ and $\dot{R}$.  By a standard Legendre
transformation this can be converted into a collective
Hamiltonian. This collective Hamiltonian is given by
\begin{equation}
H_{\rm rot} = M_0+ \frac{1}{2 I_1} \sum_{A=1}^3 {\hat{J}_A'}{}^2
\, + \, \frac{1}{2 I_2} \sum_{A=4}^7 {\hat{J}_A'}{}^2   \; ,
\label{collective}
\end{equation}
where $M_0$ is the mass of the static soliton and $I_1$ ($I_2$) is
the moment of inertia within (out of) the SU(2) subspace, and
$\hat{J}_A'$ are generators of SU(3) in a body-fixed
(co-rotating) frame.  The moments of inertia are computed in the
standard way:
\begin{eqnarray}
\hat{J}_A' &=& I_1 \dot{\theta}_A' \; \; \; {\rm for} \; \; \; A=1,2,3
\nonumber\\
 \hat{J}_A' &=& I_2 \dot{\theta}_A' \; \; \; {\rm for} \; \; \;
A=4,5,6,7\; .
\end{eqnarray}
 The numerical values of the moments of inertia are
model dependent but the structure of the collective Hamiltonian is
not.

Before proceeding, a few comments on the generality of this
procedure is in order. While it is easy to see that the precise
procedure outlined above is only applicable to models in which
the chiral field is the only degree of freedom, the ultimate
result of any analysis valid at large $N_c$ will yield a
collective Hamiltonian precisely of the form of
eq.~(\ref{collective}) {\it provided the assumption that the
collective and vibrational modes decouple is valid}. In
particular, modifications to this procedure are needed for models
which have first time derivatives of fields in the Lagrangian.
For example, in models with explicit quark degrees of freedom,
one has a nonzero quark contribution to the moments of inertia
which can be computed via a standard ``cranking'' procedure
borrowed from many-body physics\cite{crank}, and introduced into
chiral soliton physics in refs.~\cite{CB1,CB2}.  The method was
first used for the SU(3) quark-soliton mode by McGovern and
Birse\cite{MB}. With such a treatment the quark contributions to
the moments of inertia are given by
\begin{eqnarray}
I_{1}^{\rm quark} &=& \frac{N_c}{2} \sum_{i} \frac{|\langle i |
\lambda^A | 0
\rangle|^2}{\epsilon_i -\epsilon_0} \; \; A=1,2,3 \nonumber\\
 I_{2}^{\rm quark} &=& \frac{N_c}{2} \sum_{i} \frac{|\langle i | \lambda^A | 0
\rangle|^2}{\epsilon_i -\epsilon_0} \; \; A=4,5,6,7
\label{quarkcont} \end{eqnarray} where $|i \rangle$ ($\epsilon_i$)
are the single-particle quark  eigenstates (eigenenergies) for
quarks propagating in the static background of the hedgehog
fields.  The state $i=0$ corresponds to the quark ground state;
the factor of $N_c$ reflects the fact that there are $N_c$ quarks
in the system, each in the ground state in a mean-field treatment.
However, for the present purpose the key point is that although
the quantization requires a treatment somewhat more sophisticated
than that used in eq.~(\ref{ut}), the sole effect of this added
sophistication is a change in the numerical value of the moments
of inertia; the structure of the collective Hamiltonian in
eq.~(\ref{collective}) remains valid provided the central
assumption that the collective and rotation degrees of freedom
decouple is correct.

In quantizing the collective Hamiltoninan in
eq.~(\ref{collective}), a constraint plays an essential role:
\begin{equation}
{J'_8}=-\frac{N_c B }{2\sqrt{3}} \;  ,\label{quantcond}
\end{equation}
where $B$ is the baryon number.  In the context of Skyrme-type
models this quantization condition is deduced from the topology
of the Wess-Zumino term \cite{SU3Quant}.  As noted by
Witten\cite{Wit1}, this constraint can be understood in analogy to
the constraint on the body-fixed angular momentum which arises
when quantizing a charged particle in the field of a magnetic
monopole.

The masses which emerge from eqs.~(\ref{collective}) and
(\ref{quantcond}) may be found easily.   Using the fact that
\begin{equation}
\sum_{A=1}^8 (\hat{J}'_A){}^2 = \sum_{A=1}^8 (\hat{J}_A){}^2 = C_2
\end{equation}
 where $C_2$ is the quadratic Casimir operator and $\hat{J}_A$
 is a generator in the space-fixed frame, one
can rewrite the collective Hamiltonian as
\begin{equation}
H_{\rm rot} = M_0 + \frac{1}{2 I_2} \sum_{A=1}^8 \hat{J}_A{}^2 \,
+ \, \frac{I_2-I_1}{2 I_1 I_2} \sum_{A=1}^3 \hat{J}_A'{}^2  -
\frac{1}{2 I_2} \hat{J}'_8{}^2 \; .
\end{equation}
Equation (\ref{quantcond}) can be used to replace the last term.
Moreover, the intrinsic SU(2) subspace satisfies the usual SU(2)
soliton rule that $I=J$. Together these relations allow one to
express the eigenstates of $H_{\rm rot}$, {\it i.e.,} the physical
masses:
\begin{eqnarray}
M & = &  M_0 + \frac{C_2}{2 I_2} + \frac{(I_2 -
I_1) J (J+1) }{2 I_1 I_2} - \frac{N_c^2}{24 I_2} \; , \nonumber \label{mass} \\
{\rm with} \; \; &C_2 & =  \left( p^2 + q^2 + p q + 3(p +q)\right
)/3  \; ,
 \end{eqnarray}
where  $C_2$,  the quadratic Casimir, is expressed  in terms of
the traditional labels $p,q$ which denote the SU(3)
representation. The quantization condition in
eq.~(\ref{quantcond}) greatly limits the possible SU(3)
representations which can be associated with physical states:
those SU(3) representations which do not contain states with
hypercharge equal to $N_c /3$ are clearly unphysical: if the
hypercharge in a body-fixed frame satisfies
eq.~(\ref{quantcond}), then that representation will of necessity
include a state with that hypercharge. Angular momentum also
limits the physically allowed representations. In the body-fixed
frame the SU(2) manifold has  $I=J$ and $S=0$, which implies that
the number of angular momentum states associated with
representation,  (2 J+1),  must equal the number of states in the
representation with $S=0$. This whole procedure is rigid-rotor
collective quantization. The moments of inertia are treated as
constants independent of the rotational state of the system and
in that sense corresponds to a rigid rotor.

  There is a practical issue
about how one one chooses to implement this procedure. One natural
approach would be to choose to quantize the theory at large $N_c$
and then to treat systematically all $1/N_c$ corrections.   An
alternative approach would be to fix $N_c=3$ at the outset when
implementing the quantization condition of
eq.~(\ref{quantcond}).  If the approach is valid and if $N_c=3$
can be considered large, it ought not make any difference which
of these approaches is used. The choice of taking $N_c=3$ at the
outset has been the one typically made\cite{SU3Quant}. Making
this choice, it is straightforward to see that the lowest-lying
states  are:
\begin{eqnarray}
J=1/2 \; \; \; (p,q) &=& (1,1) \; \; \;({\rm octet})  \nonumber \\
J=3/2 \; \; \; (p,q) &=& (3,0) \; \; \;({\rm decuplet})  \nonumber \\
J=1/2 \; \; \; (p,q) &=& (0,3) \; \; \;({\rm anti-decuplet}) \;
.\label{multi}
\end{eqnarray}
 Equation (\ref{mass}) can be used to find the mass splitting
 of the decuplet and the anti-decuplet relative to the octet:
\begin{eqnarray}
M_{10} - M_{8} & = & \frac{3}{2 I_1} \; ,\label{10-8} \\
M_{\overline{10}} - M_{8} &  = & \frac{3}{2 I_2} \;  .
\label{10bar-8}
\end{eqnarray}

The prediction of an anti-decuplet representation is at the heart
of the issue.  The anti-decuplet contains a state with s=+1
(which has been identified with the $\theta^+$).  Such a state is
necessarily exotic, even in the large $N_c$ limit.

In outlining the rigid-rotor quantization procedure, large $N_c$
QCD considerations appeared explicitly only when discussing the
quantization constraint of eq.~(\ref{quantcond}).  In fact, large
$N_c$ considerations are at the core of the method and have been
used implicitly throughout in two essential ways.  In the first
place, large $N_c$ is necessary for the justification of the
classical static hedgehog configurations in an underlying quantum
theory.  Standard large $N_c$ scaling rules for couplings ensure
that effects of quantum fluctuations around the hedgehogs are
suppressed by $1/N_c$. Large $N_c$ also plays a central role in
justifying the semi-classical treatment in the rigid rotor
collective quantization which requires the decoupling of the
collective motion from the vibrational motion around the static
of the hedgehog. This is also suppressed by $1/N_c$, at least in
certain situations. It should be clear from the previous comment,
however, that the validity of the rigid-rotor collective approach
depends on restricting its application to those modes which
decouple from the vibrational ones.  This issue is at the heart
of the present paper.

Most treatments of SU(3) solitons identify the octet and decuplet
states with the known $N_c=3$ octets and decuplets.  Until fairly
recently, the anti-decuplet was often assumed to be a large
artifact of large $N_c$ QCD and hence ignored for essentially the
same reason that I=J=5/2 baryons are ignored in SU(2) soliton
models \cite{ANW}. The central point of ref.~\cite{DiaPetPoly} is
that the anti-decuplet should be taken seriously. The authors of
ref.~\cite{DiaPetPoly} distinguish the situation of anti-decuplet
for SU(3) solitons from the J=I=5/2 baryons in SU(2) in terms of
their widths. The J=I=5/2 baryon width is predicted to be so wide
with real world parameters that the state can not be
observed\cite{CohGri}, while the computed $\theta^+$ width turns
out to be quite small.

\section{Large $N_c$ Scaling of Mass Splittings \label{LargeN}}

The analysis presented above is based on making the choice to fix
$N_c=3$ at the outset when implementing the constraint of
eq.~(\ref{quantcond}).  This is not a reasonable way of doing
phenomenology but may obscure the large $N_c$ scaling of the
system.  Consider the scaling of the splitting as given in
eq.~(\ref{10bar-8}).  Both $I_1\sim N_c$ and $I_2 \sim N_c$,
eq.~(\ref{10bar-8}) and this appears to imply that $(M_{10} -
M_{8}) \sim 1/N_c$ and $(M_{\overline{10}} - M_{8}) \sim 1/N_c$.
Thus the exotic states appear to behave similarly with the
non-exotic states at large $N_c$. However, this is misleading. It
has been known for some time that the exotic states have mass
splittings relative to the ground state of order $N_c^0$ and not
$N_c^{-1}$\cite{KlebKap,Coh03,DP}.  To see how this arises we
explore the implementation of eqs.~(\ref{quantcond}) and
(\ref{mass}) for $N_c$ arbitrary and large.

In studying large $N_c$ baryons it is useful to restrict attention
to the case of odd $N_c$; this ensures that the baryons are
fermions. The lowest-lying representation for odd $N_c$
consistent with the quantization condition in
eq.~(\ref{quantcond}) is $ \left (p,q \right ) = \left( 1,
\frac{N_c-1}{2} \right) $; this representation can easily be
shown to have $J=1/2$ using the method described in
sect.~{\ref{Rigid}.  The Young tableau  for this representation
is given in diagram a) of fig.~\ref{young}. Note that this
representation does not correspond to {\it any} of the usual
representations at $N_c=3$, and, in particular, is not an octet.
However, the states in this representation do include those in
the usual octet.  Accordingly it is natural to take this
representation to be the large $N_c$ generalization of the octet.
This representation may be denoted ``8'' ; the quotation marks
act as reminders that this is not the octet but its large $N_c$
generalization.

Similarly, the next representation has $ \left ( p,q \right ) =
\left( 3,\frac{N_c-3}{2} \right ) $ which has $J=3/2$.  The Young
tableau for this representation is given in diagram b) of
fig.~\ref{young}. This representation contains all the states in
the usual decuplet and can be regarded as the large $N_c$
generalization of the decuplet; accordingly, this representation
is denoted by ``10''. The mass relation in eq.~(\ref{mass}) then
gives the mass splitting of the ``10'' from the ``8'':
\begin{equation}
M_{ ``10"} - M_{``8"}  = \frac{3}{2 I_1}
\label{108quote}
\end{equation}
The splitting obtained at large $N_c$ is thereby identical to the
analogous result for the decuplet-octet splitting in
eq.~(\ref{10-8}) which was obtained with the assumption $N_c=3$.
The $N_c$ scaling of this splitting is found to scale as
$N_c^{-1}$ (since $I_1$ scales as $N_c$). Thus the representations
become degenerate as $N_c$ goes to infinity. This is for a deep
reason: all of these non-exotic collective states are part of one
contracted SU(2 $N_f$) representation which emerges at large
$N_c$ (as will be discussed in subsection \ref{cons}).

\begin{figure}
\includegraphics{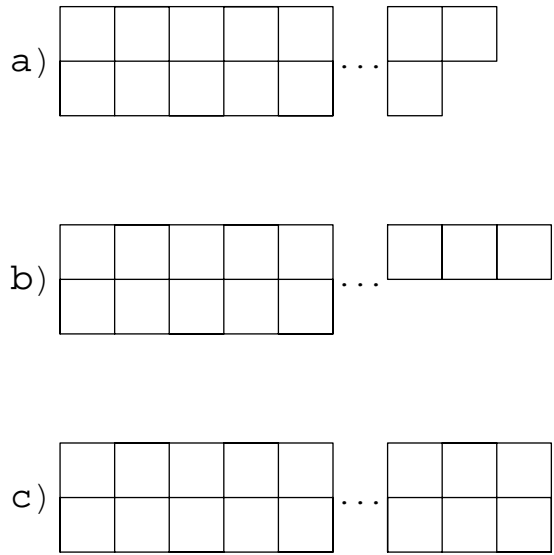}
\caption{ Young tableau for arbitrary but large $N_c$: a) the
``8'' representation with $(p,q)=\left( 1,\frac{N_c-1}{2} \right
)$; b)the ``10'' representation with $(p,q)=\left(
3,\frac{N_c-3}{2} \right )$: c) the ``$\overline{10}$''
representation with $(p,q)=\left( 0,\frac{N_c+3}{2} \right )$. The
Young tableau in a) and b) have $N_c$ boxes;  the tableau in c)
has $N_c+3$ boxes   } \label{young}
\end{figure}

To study of exotic states we need the large $N_c$ generalization
of  the $\overline{10}$ representation. The key feature of the
$\overline{10}$ is that it is the lowest-lying representation
that contains a state with strangeness +1. Accordingly, its large
$N_c$ analog should be the lowest-lying representation that
includes an exotic state with strangeness +1.  This is readily
seen to be $ \left ( p,q \right ) = \left( 0, \frac{N_c+3}{2}
\right ) $ and has $J=1/2$. This representation is associated
with the Young tableau c) in fig.~\ref{young} and is denoted as
``$\overline{10}$''. The excitation energy of this representation
is obtained via eq.~(\ref{mass}):
\begin{equation}
 M_{``\overline{10}"} - M_{``8"}  = \frac{3 + N_c}{4 I_2} \; .
 \label{10bar8quote}
\end{equation}

By construction,  eq.~(\ref{10bar8quote}) agrees with
eq.~(\ref{10bar-8}) when $N_c=3$.  However, there is an explicit
$N_c$ in the numerator of the right-hand side while the
denominator is proportional to $I_2$ which scales as $N_c$. Thus,
 at large $N_c$, the scaling is given by
\begin{equation}
 M_{``\overline{10}"} - M_{``8"} \sim N_c^0 \; \; .
 \label{scaling}
\end{equation}
In the large $N_c$ limit the ``$\overline{10}$''  does not become
degenerate with the ``8''.  It is easy to see that this behavior
is generic for exotic representations.  That is, any
representation which contains at least one manifestly exotic
state will have a splitting from the ground state which is finite
as $N_c \rightarrow \infty$.  It is interesting to note that this
behavior is characteristic of typical vibrational
excitations\cite{Wit3}.

The study of the excitation  energies as calculated via
rigid-rotor quantization reveals a fundamental difference between
exotic and non-exotic representations and the same fundamental
difference is also seen via large $N_c$ consistency rules. The
non-exotic representations become degenerate with the ground
state at large $N_c$ while the exotic representations do
not---they remain split from the ground state at leading order in
the $1/N_c$ expansion.

\section{Rigid-rotor quantization versus large $N_c$ QCD\label{Args}}

This section demonstrates that the predictions of rigid-rotor
quantization do not appear to follow from the known behaviors of
such states in large $N_c$ QCD. This is seen from a wide variety
of perspectives. Five rather general arguments are given all of
which imply that the results of rigid-rotor quantization do not
follow with the known behavior of large $N_c$ QCD.

\subsection{Exotic baryon widths \label{width}}

This subsection focuses on the $N_c$ scaling of the width of the
exotic resonance. The essence of this argument is quite simple and
requires two things be demonstrated. The first is that rigid-rotor
quantization, if valid, must predict widths which vanish at large
$N_c$; the second, that the width of exotic states with
rigid-rotor quantization is, in fact, of order $N_c^0$ implying
that it is invalid. Before demonstrating both of these points, it
should be noted that this argument is {\it formal}.  The issue of
relevance is not whether the width of the $\theta^+$ is {\it
numerically} large or small, but how it scales with $N_c$.  The
key point is that if the width does not vanish as $N_c \rightarrow
\infty$, as a matter of principle, the quantization prescription
is presumably wrong.

Let us begin with the first point.  As noted in the introduction,
the question of whether the rigid-rotor quantization is valid at
large $N_c$ has a clear formal definition.  It is valid, if and
only if, the properties computed via rigid-rotor quantization
become exact as $N_c \rightarrow \infty$.  The excitation energy
of the exotic state at large in rigid-rotor quantization is given
by eq.~(\ref{10bar8quote});  the issue is whether this becomes
exact as $N_c \rightarrow \infty$.  It is not exact if that there
is a non-zero width in the large $N_c$ limit. This can be
understood in two complementary ways. One way is simply to note
that unless the width  goes to zero at large $N_c$, the state is
not well defined at large $N_c$---it does not correspond to an
asymptotic state of the theory; it has no well-defined mass and
eq.~(\ref{10bar8quote}) does not give the exact value for the
mass at large $N_c$.  An complimentary perspective is to assign
the width to be an imaginary part of the mass. However, if the
width remains finite at large $N_c$, then the mass has a
non-vanishing imaginary part at large $N_c$, in which case
eq.~(\ref{10bar8quote}) does not give the exact result for the
mass.  From either perspective, eq.~(\ref{10bar8quote}) is not
exact at large $N_c$ unless the width vanishes.

Note, for comaprison that for non-exotic collective baryons such
as the $\Delta$ , the widths do go to zero at large $N_c$ and
therefore the prediction of eq.~(\ref{108quote}) can indeed
become exact. To see this, one simply notes that the non-exotic
collective excitation energies as given by eq.~(\ref{108quote})
scale as $1/N_c$ while meson masses scale as $N_c^0$. Thus, at
large $N_c$ non-exotic collective space have no phase-space for
decay and, hence, have zero width.

Let us now turn to the case of the exotic collective baryons
which is quite different.  Recently, Praszalowicz demonstrated
that the width of the $\theta^+$ is of order $N_c^0$
\cite{Pras}.  His result is summarized below.

 Equation (\ref{scaling})
implies that the excitation energy of the ``$\overline{10}$''
representation does not vanish at large $N_c$. Thus, in general
phase space it does not inhibit decay.  Of course, phase space is
not the only reason the width can go to zero.  It is also
logically possible that the coupling between the initial exotic
baryon and the final state of meson plus baryon could vanish at
large $N_c$.  The issue boils down to whether or not this happens.

As shown by Praszalowicz \cite{Pras}, in the context of
rigid-rotor quantization, there are two possible structures for
the coupling of collective baryon states to mesons which
contribute in leading order to the width:
\begin{equation}
\hat{O}_\kappa  =  -i \frac{3}{2 M_B} \left (G_0
\hat{O}^{0}_{\kappa j} + G_1 \hat{O}^{1}_{\kappa j} \right ) p_j
\label{cupop}
\end{equation}
with
\begin{equation}
G_0 \hat{O}^{0}_{\kappa j} = D^{8}_{\kappa j} \; \; \;  {\rm and}
\; \; \; G_1 \hat{O}^{0}_{\kappa j} = d_{j k l} D^{8}_{\kappa k}
\hat{S}_l
\end{equation}
where $\kappa$ indicates the meson species and $\hat{S}$ is the
spin operator.  While the values of the coupling constants, $G_0$
and $G_1$, are model dependent, their $N_c$ scaling is known:
\begin{equation}
G_0 \sim N_c^{3/2} \; \; \; G_1 \sim N_c^{1/2} \; .\label{Gscale}
\end{equation}
Using the collective SU(3) wave functions for the ``8'' and
``$\overline{10}$'' representations one finds:
\begin{widetext}
\begin{equation}
\left |\langle \theta^+,s=\downarrow|\hat{O}_{K 3}|N,
s=\downarrow \rangle \right |^2 = \frac{3}{\left (M_N +
M_{\theta^+} \right )^2} \frac{3 (N_c+1)}{(N_c+3)(N_c+7)} \left
[G_0 - \frac{N_c+1}{4} G_1 \right ]^2 p^2. \label{MEsq}
\end{equation}
\end{widetext}

The $G_1$ term enhanced by a factor of $N_c$ (coming from the
integral over the collective wave function) compensates for the
fact that it is characteristically smaller than $G_0$ by a factor
of $N_c$. Using the scaling in eq.~(\ref{Gscale}) and the fact
the phase space implies $p \sim N_c^0$, one sees the matrix
element is of order $N_c^0$.  Combining this with the phase space
one sees that
\begin{equation}
\Gamma_{\theta^+ }\sim N_c^0 \label{widtheq}.
\end{equation}

The scaling in eq.~(\ref{widtheq}) completes the demonstration.
The width as computed using states from rigid-rotor quantization
shows that the rigid rotor mass formula of eq.~(\ref{10bar8quote})
does not correctly give the mass at large $N_c$: rigid-rotor
quantization does not appear to be self-consistent in large $N_c$
QCD .

\subsection{Time scales for collective and vibrational motion \label{tscales}}

The key to the success of collective quantization is a clear
separation of time scales between the collective motion and the
intrinsic motion.  As a general principle, to quantize the theory
one must first enumerate the degrees of freedom. The easiest way
to do this is via the study of small classical fluctuations around
the classical soliton: each classical mode  associated with small
amplitude fluctuations is a degree of freedom which can be
quantized.  In studying these small amplitude fluctuations one
finds that although the typical modes have a vibrational frequency
which scales as $N_c^0$, of necessity there will also be zero
frequency modes. Such zero modes arise because symmetries
associated with the underlying theory are broken by the soliton
configurations.

Once these modes have been identified at the classical level,
they can  be canonically quantized.  As a practical matter, a
parametric expansion is used as a systematic way to organize
things\cite{sol}.   It can be shown self-consistently that for
properties of low-lying states the neglect of these couplings
leads to a $1/N_c$ correction to the soliton mass as compared to
a leading order value which goes as $N_c$, and a correction due
to the quantized zero modes which goes as $N_c^0$.  The
quantization of the collective degrees of freedom associated with
the zero modes can be handled separately for exactly the same
reason---the effects of the coupling of the collective to the
vibrational degrees of freedom on the soliton mass is suppressed
by $1/N_c$.

The collective degrees of freedom  are those which at the
classical level correspond to zero modes;  modes having a finite
frequency are quantized as vibrations.  As will be discussed in
sect.~\ref{revisit}, the modes associated with the exotic
excitations turn out not to be collective in this sense; they are
associated with modes which have finite frequencies at small
amplitude and, thus, act as vibrations.

The effect of the quantization of the collective degrees of
freedom on the soliton mass is typically of order $1/N_c$. At the
classical level the degrees of freedom are flat and have no
natural time scale---the classical motion associated with them
can occur arbitrarily slowly.  The quantization of these
collective modes keeps them slow---the typical time scales are of
order $N_c$.   The characteristic angular speed for non-exotic
collective coordinates is $1/N_c$---it goes as the inverse of the
moment of inertia since the associated angular momentum is
typically of order unity and the motion is of large amplitude
(order $N_c^0$ typically circumnavigating the collective space),
so the characteristic time, $\tau$, is of order $ N_c$. In
semi-classical motion one has a general energy-time uncertainty
relation:
\begin{equation}
\Delta E \sim 1/\tau  \; \label{tau},
\end{equation}
where $\Delta E$ is the typical splitting between levels.  This
naturally gives energy splittings of order $1/N_c$.   It is
straightforward to see that this behavior reproduces what is seen
in eq.~(\ref{108quote}).

The time scale for the non-exotic collective motion ($N_c$) scales
in a qualitative different way from that of the internal
vibrational motion (shown below to scale as $N_c^0$). This
difference is critical for the success of the quantization
program, particularly for the separate treatment of the
collective and vibrational degrees of freedom. Note that there
exist vibrational modes with the same quantum numbers  as those
of the collective degrees of freedom.  One generally expects
degrees of freedom with identical quantum numbers to couple.
However, two degrees of freedom with identical quantum numbers
naturally remain weakly coupled if the characteristic times of
the two degrees of freedom are widely separated.  In this case
there is a Born-Oppenheimer type separation of the dynamics of
the modes. For the case of collective rotational degrees of
freedom this is precisely what happens and the modes are
effectively decoupled for low-lying states.

Let us now turn to the behavior of the exotic states.  It is
apparent from eq.~(\ref{10bar8quote}) that  excitations with
exotic quantum numbers have excitation energies of order $N_c^0$,
implying that even for the lowest-lying exotic state, $\Delta E
\sim N_c^0$. This in turn implies that $\tau \sim N_c^0$, where
$\tau$ is the characteristic time scale of the motion.  This is
highly problematic from the viewpoint of collective motion. In the
first place, this is manifestly inconsistent with the motion
being collective in the sense it corresponds to large amplitude
motion.  Equation (\ref{collective}) implies that that excitation
energies of order $N_c^0$ correspond to motion with ${J'_A}^2/I_2
\sim N_c^0$ (with A=4,5,6,7). This, along with the fact that
$I_c\sim N_c$, further implies $J'_A \sim N_c^{1/2}$, and the
relevant angular velocities are of order $N_c^{-1/2}$. Therefore
the motion is slow---the angular velocity goes to zero as $N_c$
goes to infinity.  However, we also know that the characteristic
time for this motion is of order $N_c^0$. Clearly this is not
possible unless the typical angular displacement associated with
this motion is of order $N_c^{-1/2}$.  Thus, this motion is
confined to a region of angular space which goes to zero at large
$N_c$: the motion does not explore the full collective space.  In
this sense the motion is clearly not collective. This should be
contrasted to the case of low-lying non-exotic states which have a
characteristic time scale of $N_c^1$, a characteristic angular
velocity of $N_c^{-1}$ and, hence, a characteristic angular
displacement of $N_c^0$, which is of large amplitude and
subsequently collective.

Given the fact that motion is not collective,  a collective
quantization as in the rigid-rotor approach is presumably invalid.
Moreover, it is easy to understand why it can fail. The time scale
for even the lowest-lying exotic states, is of order $N_c^0$; this
is the as for vibrational motion. In the absence of such a scale
separation, there is nothing to prevent the ostensibly collective
mode from mixing strongly with vibrational motion with the same
quantum numbers and {\it a priori} that is exactly what one
expects to happen. When this occurs the rigid-rotor approximation
based on the separate dynamical treatment of the two types of
motion clearly fails.

\subsection{Predictions from large $N_c$ consistency rules\label{cons}}

The introduction stressed the remarkable fact that the prediction
of the mass in refs.~~\cite{Pres,DiaPetPoly,WalKop,BorFabKob,Kim}
was independent of dynamical details of the model. The key model
independent relations were that the SU(3) symmetry breaking used
for the exotic states was identical to that used for the ground
state band. It was largely because of this model independence that
the prediction should be taken seriously.  Model-independent
predictions are special; the very fact that they do not depend on
model details suggests that they may well be reflecting the
underlying structure of large $N_c$ QCD.

Of course, just because a particular relation obtained in a chiral
soliton model does not depend on the detailed dynamics of the
model, need not imply the result is truly model independent
results of large $N_c$ QCD.   There is strong evidence, however,
these relations are, in fact, truly model independent. {\it All}
of these relations seen to date ({\it with the exception of those
involving exotic baryons computed via rigid-rotor quantization}),
are known to be correct model-independent predictions of large
$N_c$ QCD\cite{GS,DM,Jenk,DJM}. These include relations of
standard static observables (such as magnetic moments or axial
couplings) as considered in ref.~\cite{AdkNap}, as well as the
rather esoteric quantities, such as the non-analytic quark mass
dependence of observables near the chiral limit\cite{nc}, or
relations between meson-baryon scattering observables in
different spin and flavor channels \cite{MatPes,CohLeb}.  All of
these can similarly be derived in a ``model-independent'' manner
in the chiral soliton model and are also derivable by a truly
model-independent way.

The basis for demonstrating these model-independent predictions is
the use of large $N_c$ consistency rules\cite{GS,DM,Jenk,DJM}:
the large $N_c$ predictions for various related quantities are
not self consistent unless various relations are imposed.  For
example, according to Witten's large $N_c$ counting rules
generically meson-baryon coupling constants scale as $N_c^{1/2}$
\cite{Wit3}. Thus, two insertions of coupling constants will yield
a contribution to meson-baryon scattering of order $N_c^1$, but
unitarity ) requires the scattering to be of order $N_c^0$. This
can only be satisfied if the baryons form towers of states which
are degenerate at large $N_c$ with the values of the coupling
constants related to one another by geometrical factors (up to
$1/N_c$ corrections)\cite{GS,DM,Jenk,DJM}.  These factors turn
out to be precisely the ones found in the chiral soliton models.
Other quantities can be derived via similar means.

The results of this type of analysis are well known: A contracted
SU(2$N_f$) symmetry emerges in the large $N_c$ limit. Baryon
states fall into multiplets of this contracted SU(2$N_f$), and the
low-lying states in these multiplets are split from the ground
state by energies of order $1/N_c$---these excitations with the
SU(2$N_f$) multiplets are collective.  In the space of these
collective states all operators can ultimately be expressed in
terms of generators of the group and from this, relations can be
obtained.

The key issue here is simply that the multiplet of low-lying
baryons has been explicitly constructed---it coincides exactly
with the low spin states of a quark model with $N_c$ quarks
confined to a single s-wave orbital\cite{DJM}. It is well known
that at large $N_c$ there are no low-lying collective baryon
states ({\it i.e,} states with excitation energies of order
$1/N_c$})  with exotic quantum numbers.  This neatly mirrors the
analysis of sect.~\ref{LargeN}: exotic states have excitation
energies of order $N_c^0$.

This situation is quite problematic, however, if one wishes to
assert that relations for the exotic states are truly model
independent.  The difficulty is simply that the exotic states are
not in the same contracted SU(2$N_f$) multiplet as the ground
band baryons.  This means that group theory alone cannot relate
matrix elements in the ground band to matrix elements involving
exotic states: the standard large $N_c$ consistency rules do not
allow one to relate any property of the exotic states to the
ground band states.

At a minimum this implies that the apparently model-independent
predictions of exotic state properties seen from rigid-rotor
quantization have not been shown to be truly model independent.
This is characteristically different from the properties of
non-exotic states which are related to one another by large $N_c$
consistency rules. Of course, just because the large $N_c$
consistency conditions do not give the relations seen for exotic
states in rigid rotor computation does not by itself mean that
these relations are wrong; it merely means we do not know them to
be correct. However, it does mean that the principle reason to
take the predictions of the chiral soliton model seriously---their
apparently model-independent status---has no known basis in QCD.

\subsection{Solitons for large $N_c$ in a world with heavy quarks}

The predictions in refs.~\cite{DiaPetPoly} were based on chiral
soliton models.  However, this is somewhat misleading.  The basic
arguments of rigid-rotor quantization  depends critically on the
SU(3) flavor symmetry of the underlying theory and the fact that
the classical solution breaks both rotational and flavor
symmetries in a correlated way resulting in a hedgehog
configuration. However chiral symmetry  plays role only through
the chiral anomalies encoded in the Wess-Zumino term which leads
to the constraint in eq.~(\ref{quantcond}).  However, as shown in
appendix \ref{const}, eq.~(\ref{quantcond}) can be derived
directly at the quark level with no reference to anomalies. Thus,
the logic underlying collective quantization applies to all SU(3)
symmetric theories which have hedgehog mean-field solutions. In
particular, it applies to models of QCD with three degenerate
flavors in the limit where all quark masses are heavy; $m_u = m_d
= m_d \equiv m_q$ with $m_q>> \Lambda$ where $\Lambda$ is the QCD
scale (provided that such models produce a hedgehog at the
mean-field level).  In this section it is shown that the
application of the rigid-rotor quantization in such a regime
gives results which are manifestly wrong suggesting that something
is wrong with the underlying logic.

In fact, in this regime one need not consider a model of QCD, but
rather QCD itself.  Recall that in Witten's original derivation
of baryon properties in large $N_c$ QCD the case of heavy quarks
was considered for simplicity (it was later argued that the
conclusions are valid for the case of light quarks)\cite{Wit3}.
The derivation is based on the fact that in this limit the quarks
are non-relativistic and can be described via the many-body
potentials arising from gluon exchange.  Witten then demonstrated
that the Hartree mean-field approximation becomes valid in the
large $N_c$ limit, that all $N_c$ quarks are in the same single
particle wave function  (modulo their color degree of freedom)
and that the size of this orbital is independent of $N_c$.  In
doing this analysis, the role of flavor and spin degrees of
freedom was not highlighted.

Consider the role played by spin and flavor in this system. From
the analysis of Dashen, Jenkins and Manohar we know that for
states with $J,I \sim N_c^0$ at leading order in the $1/N_c$
expansion the only interactions which contribute are either spin
and isospin independent or both spin dependent and isospin
dependent\cite{DJM}. The spin dependence enters solely through the
magnetic coupling of the gluons to the quarks.  Note, however,
that the underlying magnetic interaction is small for large quark
masses since the quark magnetic moment goes as $1/m_q$.  Thus all
spin-flavor dependent interactions are small in the combined
large $N_c$ and heavy quark limits.

 The heavy quark limit implies
that the spatial shape of the single-particle levels does not
depend on  spin and flavor.  Assuming that rotational symmetry is
not broken at this order, the orbitals will be s-wave. However,
the Hartree state is highly degenerate at this order: any spin and
flavor orientation is equivalent.  If one one includes the
leading $1/m_q$ correction, this degeneracy is broken (but at this
order the spatial part of the wave functions are unchanged) and
one simply chooses the single-particle state to have a spin-flavor
orientation which minimizes the spin-flavor interaction at order
$1/m_q$.  The precise form of this spin-flavor orientation depends
on the sign of the interaction which either favors a ground of
the baryon with maximal spin or minimal spin.  If it favors
maximal spin, then all quarks will have the same well-defined spin
and flavor projections. In contrast, if it favors minimal spin
(as is believed to happen in nature, which will be assumed here)
the spin-flavor part of the state takes the conventional hedgehog
form:
\begin{equation}
|h \rangle = 2^{-1/2} \left ( |\uparrow d \rangle - |\downarrow u
\rangle \right ) \label{h}
\end{equation}
where the arrows indicate spin projection and the letters
indicate flavor. It  is assumed for simplicity that the hedgehog
is in the u-d subspace with grand-spin of zero.

It is easy to see that there are low-lying single-particle excited
states for quarks propagating in the background of the Hartree
potential generated by the hedgehog state. There are six distinct
spin-flavor states, one of which, the state, $|h\rangle$, is, by
construction, the lowest-lying state for the Hartree potential.
These states differ from the lowest lying state in energy by an
amount of order $\Lambda^2/m_q$, where the $1/m_q$ is due to the
spin dependence as noted above and the factors of $\Lambda^2$
follow from dimensional analysis.  Furthermore, all flavor
generators (except for $\lambda_8$), when acting on the hedgehog,
will produce a superposition of these low-lying excited
single-particle states. Thus, if one computes the moments of
inertia using eqs.~(\ref{quarkcont})  one has the following
scaling in $N_c$ and $m_q$:
\begin{equation}
I_1^{-1} \sim \frac{N_c \Lambda^2}{m_q} \; \; I_2^{-1} \sim
\frac{N_c \Lambda^2}{m_q} \; . \label{hqmi}
\end{equation}

Equation (\ref{hqmi}) has profound implications for excitation
energies assuming that rigid-rotor quantization is legitimate.
Consider first the behavior of the non-exotic states, such as
those in the ``10'' representation.  Equations (\ref{108quote})
along with eq.~(\ref{hqmi}) imply that scaling of the excitation
energy with $m_q$ and $N_c$ goes as
\begin{equation}
M_{``10"}-M_{``8"} \sim \frac{\Lambda^2}{m_q N_c} \; .
\end{equation}
This behavior is precisely what one expects.  It is suppressed by
$1/N_c$ since it involves the excitation of a single quark and it
goes as $1/m_q$ since it relies on a magnetic gluon-quark
coupling.

In a similar way one can compute the scaling of the excitation
energy of the exotic states such as the ``$\overline{10}$''.
Equations (\ref{10bar8quote}) and eq.~(\ref{hqmi}) imply that
this excitation scales as
\begin{equation}
M_{``\overline{10}"}-M_{``8"} \sim \frac{\Lambda^2}{m_q} \; .
\label{es}\end{equation} However, this scaling is quite
problematic on physical grounds. To construct an exotic state
such as the ``$\overline{10}$'' one must include an extra
quark-antiquark pair relative to the ``8''. Recall that by
assumption the quarks are heavy enough to be non-relativistic;
the binding energies of the quarks are much smaller than the
mass.  The excitation energy of the exotic state must then be $2
m_q$ plus small corrections; it grows with $m_q$. This is in
contradiction with eq.~(\ref{es}) which has the excitation energy
decreasing with increasing $m_q$ assuming that rigid-rotor
quantization is valid.  The contradiction implies that the
assumption is false: rigid-rotor quantization is not valid for the
exotic states.  As noted above, the logic underlying rigid-rotor
quantization does not depend on the quark mass and the failure for
the present case indicates that the logic is flawed.

As pointed out recently by Pobyltisa\cite{Pob}, the issue is even
more stark if presented in the context of an SU(3) symmetric
non-relativistic quark model.  In this case the model space can be
constructed to have only quark---and no anti-quark---degrees of
freedom.  However, the interactions in such a theory can be
chosen to reproduce the same $N_c$ scaling seen in QCD. Again, as
$N_c$ becomes large the Hartree approximation becomes
increasingly well justified and again the Hartree minimum will be
of a hedgehog configuration (provided that the ground state
orbitals turn out to be s-waves).  The justification for doing
rigid-rotor quantization is identical to that in the chiral
soliton case. In such a model both $I_1$ and $I_2$ can be
computed using the standard expressions in eq.~(\ref{quarkcont})
and rigid-rotor quantization can then be implemented.  Such a
quantization implies the existence of exotic states with an
excitation energy of order $N_c$, {\it yet by construction such
states are not in the Hilbert space from which the model was
constructed}. The rigid-rotor quantization is wrong for exotic
states in such models. Of course, one might argue that such a
model does not represent QCD in a particularly realistic way.
However, that should be irrelevant to the central issue of the
justification of rigid-rotor quantization; the derivation of this
approach has always been based on general considerations and not
on the detailed structure of QCD.

\subsection{The Callan-Klebanov approach at zero SU(3) breaking}

The discussion in this paper has focused on states as computed in
the limit of zero SU(3) symmetry breaking.  As was argued above,
the fundamental issues of principle interest here can be most
cleanly addressed if detailed numerical questions, such as whether
the SU(3) breaking effects are too large to justify perturbation
theory, aside. One powerful way to focus on the problem is to
consider how to treat the problem in the presence of SU(3)
symmetry breaking and then consider the limit as this breaking
goes to zero.

The basic formalism of how to treat chiral solitons in the
presence of SU(3) breaking was developed long ago by Callan and
Klebanov\cite{CK}.  The logic underlying this approach is very
simple.  If $m_s - m_q$ is greater than zero (where $m_q$ is the
light quark mass), then at the classical level the minimum energy
configuration is a hedgehog in the u-d subspace.  A flavor
rotation in the 4,5,6 or 7 directions will yield a classical
configuration with higher energy.  Thus these directions are not
flat collective ones but are unambiguously of a vibrational
character.  The formalism is based on the straightforward
quantization of these vibrational degrees of freedom along with
the collective quantization of the SU(2) degrees of freedom in
the presence of the Wess-Zumino term. It is often called the
``bound state approach'' since states with negative strangeness
are viewed as a bound state of an SU(2) skyrmion with an
anti-kaon.  Here it will be referred to as the Callan-Klebanov
approach in order to avoid confusion when discussing states of
positive strangeness which are unbound. It should be noted that a
principal reason why this approach was introduced was to deal with
circumstances where SU(3) symmetry breaking was large enough so
that simple perturbative treatments were potentially unreliable.
But, as stressed in ref.~\cite{IKOR}, the formalism should be
valid regardless of the size of the symmetry breaking and, in
particular, holds as $ m_s - m_q \rightarrow 0$.

The key issue here was discussed in the recent paper by
IKOR\cite{IKOR}. Consider the Callan-Klebanov formalism as $ m_s -
m_q \rightarrow 0$.  For states with non-exotic quantum numbers it
is possible to show analytically that as this limit is approached,
the excitation energy of the bound state  goes to zero (as it is
in rigid-rotor quantization) and the structure of the vibrational
mode goes over precisely to the spatial distribution seen in the
collective moment of inertia \cite{CK,IKOR}.  This supports the
view that the Callan-Klebanov formalism applies regardless of the
size of SU(3) symmetry breaking and remains valid to the exact
SU(3) limit, and that rigid-rotor quantization is valid for
non-exotic collective states.

However, the situation is radically different in the case of
exotic s=+1 excitations.  In that case, there is no analytic
demonstration that the results of the Callan-Klebanov treatment
goes over to those of  rigid-rotor quantization as the SU(3)
limit is approached.  Of course, it is logically possible that
the two approaches are, in fact, equivalent in the SU(3) limit
but that a mathematical demonstration of this equivalence has not
yet been found. However, there is strong numerical evidence in
ref.~\cite{IKOR} that this is not the case.  In particular, for
small values of $m_K$ ({\it i.e}, for small SU(3) symmetry
breaking), if the two approaches were equivalent there ought to be
a resonant vibrational mode whose frequency is near to that
predicted in rigid-rotor quantization for all Skyrme-type models.
In fact, no such mode is seen.  Indeed, for the standard Skyrme
model, there are no exotic resonances at all.  This strongly
suggests that as the exact SU(3) limit is approached the
Callan-Klebanov approach does not become equivalent to the
rigid-rotor quantization as $ (m_s - m_q) \rightarrow 0$.

If one accepts that the standard derivation of the Callan-Klebanov
formalism is valid for this problem, then the inequivalence
between rigid-rotor quantization and the Callan-Klebanov method
for exotic states implies that the rigid-rotor quantization is not
valid. There is a mathematical subtlety associated with this: We
know the derivation of Callan and Klebanov in ref.~\cite{CK} is
valid when SU(3) breaking is large enough so that motion in the
strange direction is of a vibrational nature. As a formal matter,
this is guaranteed to happen in the large $N_c$ limit for any
finite value of  $(m_s-m_q)$.  As long as the physical quantities
of interest have a uniform limit as $N_c \rightarrow \infty$ and
$(m_s-m_q) \rightarrow 0$, then the Callan-Klebanov derivation is
automatically valid in the large $N_c$ limit for exact SU(3)
symmetry.  However, if the two limits do not commute, then taking
the SU(3) limit prior to the large $N_c$ limit (as is implicitly
done in rigid-rotor quantization ) would give results which
differ from those when taking the large $N_c$ limit first (as is
implicitly done in  the Callan-Klebanov approach).  If this is
the case, then it remains possible mathematically that both
approaches are valid in particular regimes but that their domains
of validity do not overlap.

However, on physical grounds this mathematical possibility seems
quite unlikely.  In the first place, it seems implausible {\it a
priori} that the large $N_c$ and SU(3) limits commute for
non-exotic states and then fail to exotic states.  More
importantly, there is considerable experience with cases where
the large $N_c$ limit does not commute with some other limit, and
in these cases the lack of commutativity can be traced to a clear
physical origin which is apparent at the hadronic level.  For
example, it is well known that the chiral limit and the large
$N_c$ limit do not commute for baryon quantities which diverge in
the chiral limit (such as isovector charge or magnetic radii)
\cite{nc}. In these cases the physical origin is easily traced to
the role of the $\Delta$ resonance whose excitation energy is
anomalously low parametrically (it scales as $1/N_c$) compared to
typical excited baryons and which therefore leads to a class of
infrared enhancements in loop graphs.  Thus, it is the interplay
between the two light scales in the problem---the chiral scale,
$m_\pi$ and the light scale induced at large $N_c$,
$M_\Delta-M_N$---which leads to the non-commuting behavior for
problems that are infrared singular in the combined limit. There
is nothing analogous to this when considering exotic baryons near
the SU(3) limit: the SU(3) symmetry breaking scale is the only
relevant low scale in the problem since the low-lying decuplet
type excitations play no special role.  Given these physical
arguments it is highly unlikely that the two limits do not
commute; as note above, this implies that rigid-rotor
quantization is invalid.

Apart from the physical grounds discussed in the previous
paragraph there is a clear mathematical way to understand why
numerical work of IKOR fails to find a vibrational mode whose
properties match those predicted in  rigid-rotor quantization for
exotic states.  This is the mixing between the ``collective''
mode in rigid-rotor quantization and ordinary vibrational modes
which is discussed in detail in sect.~\ref{revisit}. This mixing
occurs at leading order in both the large $N_c$ expansion and in
SU(3) breaking regardless of the ordering of limits, and thus
demonstrates mathematically that the physical arguments given
above are correct: the two limits do commute, but the rigid-rotor
quantization is not valid.

\section{Rigid-Rotor quantization Revisited \label{revisit}}
\subsection{General considerations}
The previous section provides strong evidence that rigid-rotor
quantization is not valid for the description of states with
exotic quantum numbers.  This can be seen as somewhat paradoxical
since the derivation of rigid-rotor quantization\cite{SU3Quant}
closely paralleled the derivation for SU(2) skyrmions, by Adkins,
Nappi and Witten (ANW)\cite{ANW}, which is generally agreed to be
correct. How can this method work for SU(2) solitons and for
non-exotic states in SU(3) solitons, yet fail for exotic states?

It is important to begin by recalling the fundamental assumption
made in the derivation in sect.~\ref{Rigid}; namely, that the
collective motion and the intrinsic motion are dynamically
separate.  The issue is whether this is true; as will be seen in
this section, it does not appear to be true generally for exotic
motion even at large $N_c$.

The ANW procedure amounts to putting an ansatz for a class of
allowed motion into the Lagrangian thereby obtaining a proposed
collective Lagrangian. This is done in eq.~(\ref{ut}) where the
ansatz made is that the motion corresponds to an overall
time-dependent rotation of the static soliton.  As a general
rule, the insertion of an ansatz into a Lagrangian yields a
legitimate collective Lagrangian if, and only if, all classical
solutions of the collective Lagrangian so obtained are also
solutions of the full equations of motion. Only after the
collective Lagrangian has been isolated at the classical level
can the collective Hamiltonian be found and then quantized.

As alluded to in sec. \ref{Rigid}, the  ANW treatment is strictly
valid at large $N_c$ only for models whose Lagrangians have no
first derivatives in time (such as the original Skyrme model). The
reason for this is that the ansatz in eq.~(\ref{ut}) only
corresponds to an approximate solution of the full equations of
motion in such cases. It has long been known that the method
needs to be modified for models where first derivatives in time
are present (such as in soliton models with explicit quark
degrees of freedom)\cite{CB1,CB2}.  For the SU(2) model with
explicit quarks one can find an appropriate ansatz which
corresponds to a solution of the mean-field (classical)
equations; the cranking equations provide such an ansatz. The
reason this issue becomes central here is the role of the
Wess-Zumino term. This term has an explicit time derivative, and
{\it a priori} one ought not to expect the ANW method to work
without modification.

 First consider models such as the original SU(2) Skyrme
model, which only has pion degrees of freedom and has no first
derivatives in time.  In this case it is easy to find such
families of  solutions which become exact at large
$N_c$\cite{CB2}. In particular,
\begin{equation}
U(\vec{r},t) = A e^{i \vec{\lambda} \cdot \vec{\tau} t/2}
U_0(\vec{r}) e^{-i \vec{\lambda} \cdot \vec{\tau} t /2} A^\dagger
\label{td}\end{equation} is an approximate time-dependent solution
of the classical equations of motion provided that $U_0$ is a
static solution and $\vec{\lambda}$ is an angular velocity which
is small at large $N_c$ (typically going as $N_c^{-1}$). The
parameters that specify the motion are the initial angles given in
$A$ and the angular velocities in $\vec{\lambda}$. This is an
allowable approximate time-dependent solution since the effect of
the second derivative with respect to time on the field
configuration (which is neglected when using a rotating soliton)
is of order $ \sim 1/N_c^2$ down relative to contributions to the
static solution. Thus, the neglected shifts in the fields are of
relative order $N_c^{-2}$.  This in turn implies a neglected
shift in the angular momentum of order $N_c^{-1}$ (since the
angular momentum is intrinsically of order $N_c$); this implies
that the neglected shift in the moment of inertia is of order
$N_c^0$ and may be neglected at large $N_c$ compared to the
leading order contribution of order $N_c^1$.

The ANW ansatz of eq.~(\ref{ut}) contains all the solutions of the
form of eq.~(\ref{td}). Moreover {\it all} solutions of the
classical equations of motion which emerge from the collective
Lagrangian are of this form. Thus, the ANW ansatz gives a
legitimate collective Lagrangian on the SU(2) Skyrmion. This
Lagrangian can then be quantized.

The situation is quite different if there are first-order time
derivatives.  In that case the neglected effect on the fields is
proportional to $\lambda \sim N_c^{-1/2}$ yielding a neglected
shift in the fields proportional of relative order $N_c^{-1/2}$
which in turn implies that the neglected shift in the angular
momentum is of order  $N_c^{1/2}$.  The neglected shift in the
moment of  inertia is then of order $N_c$.  These effects cannot
be neglected since the neglected contribution is of the same
order as the contribution which is kept. There is a simple way to
incorporate these effects in SU(2) solitons containing quarks. In
that case, the key point is that  one needs an ansatz for a
time-dependent solution which corresponds to the rotating
soliton.  If such a corresponding solution exists, it is
equivalent to a static solution calculated in a rotating frame
and one obtains the cranking result of eq.~(\ref{quarkcont}).

To summarize the general situation, ANW quantization, while
agreeing with the generally correct method for the case where it
was introduced, it does not directly apply to models with first
order time derivatives.  In these cases one needs to find families
of approximate classical time-dependent solutions which are
decoupled from the remaining degrees of freedom.

  In the case of
exotic motion in SU(3) solitons the Wess-Zumino term plays a
dynamical role and {\it a priori} there is no reason to believe it
should be valid in such a case. In contrast, for non-exotic motion
the Wess-Zumino term is inert (as can be seen in
eq.~(\ref{quantcond})). In principle, it is sufficient to stop the
argument here---rigid-rotor quantization has never been correctly
derived for exotic motion. The fact that the derivation was not
shown to be correct does not logically mean the result is wrong.
However, the fact that rigid-rotor quantization gives
inconsistent results as seen in the previous section indicates
that the result is, in fact, not correct.

It is useful to understand {\it why} the approach fails in a bit
more dynamical detail.  The key point is that the there is no
family of classical solutions corresponding to the exotic
collective as given by the ansatz.

Before discussing the full problem it is useful to gain some
intuition about how things work by considering a couple of
``toy'' problems.

\subsection{A charged particle in the field of a magnetic
monopole \label{mono}}
 To begin consider the following simple toy
model: the non-relativistic motion of a charged particle of mass
$m$ and charge $q$ confined on a sphere of radius $R$ with a
magnetic monopole of strength $g$ at its center. This problem was
introduced by Witten \cite{Wit1} to illustrate the role of the
Wess-Zumino term and was used to motivate the treatment of
rigid-rotor quantization for SU(3) solitons \cite{SU3Quant}.  The
monopole case is essentially similar to the Wess-Zumino case in
two fundamental aspects: a) its effect is first order in time
derivatives, and b) it is essentially topological in nature
imposing toplogical quantization rules (the Dirac condition for
monopoles, the quantization of the Wess-Zumino term for chiral
soliton  models\cite{Wit2}).  To simulate the case of SU(3)
solitons at large $N_c$ one has the following scaling rules:
\begin{equation}
q \sim N_c^0 \; \; \; \; R \sim N_c^0 \; \; \; \; g \sim N_c^1 \;
\; \; \; m \sim N_c^1 \; ;\label{toyscale} \end{equation} the
Dirac quantization condition implies that $2 q g$ is an integer.
This problem is exactly solvable.  The key physical point is that
in addition to the usual kinetic contribution to the moment of
inertia, there is an angular momentum associated with the magnetic
field energy which is given by $\vec{L}_{field} = q g \hat{r}$.
Mathematically, the key issue is the unphysical nature of a
rotation about the axis linking the charge and the monopole which
then imposes the Dirac condition and the restriction $L \ge |q
g|$. The wave functions may be expressed in terms of three Euler
angles; the independence of the the physical results on the third
angle is what gives rise to the Dirac quantization condition.

The spectrum for this problem is
\begin{equation}
E = \frac{L(L+1) - (q g)^2}{2 m R^2} \; \; {\rm with} \; \; L \ge
|q g|  \; .\label{toyspec}
\end{equation}

The analogy  with the SU(3) soliton is the following.  The
smallest allowable $L$  corresponds to the non-exotic baryons
with the (2L+1) allowable $m$'s corresponding to the different
non-exotic baryon states.  The various states with $L > q g$
correspond to exotic states.  To make this manifest we follow
Diakonov and Petrov\cite{DP} and express the energy in terms of an
``exoticness'':
\begin{equation}
E = \frac{e^2 + e (2 |q g| +1) + |q g|}{2 m R^2}  \; \; \; {\rm
with} \; \; \; e \equiv L - |q g|  \end{equation} If one focuses
on the low-lying exotic states (those states for which $e \sim
N_c^0$) which are the states of interest in hadronic physics one
finds the excitation energies are given by
\begin{equation}
E_e - E_0 = e \frac{ |q g|}{ m R^2} + {\cal O}(N_c^{-1})
\label{toyspeca}
\end{equation}
where the $N_c$ scaling is fixed from eq.~(\ref{toyscale}).

Conventional treatments of rigid-rotor quantization model their
treatment of the Wess-Zumino term on this exact treatment of this
toy problem \cite{SU3Quant}. The present purpose is different:
Here the goal is to understand  which collective degrees of
freedom can be isolated in a treatment which will ultimately be
semi-classical in nature. Of course, this toy problem has no
internal degrees of freedom---by construction it is rigid-rotor
quantization. However, it is useful to illuminate the underlying
physics of this toy problem in a manner which transparently can
be used in the case where collective and vibrational modes mix. In
particular, it is quite instructive to derive the result in
eq.~(\ref{toyspeca}) from a semi-classical treatment which can
then be generalized.

We need to first describe the classical motion of the particle.
Let us consider the particle at rest at the north pole and we can
describe all solutions of the equations of motion relative to
this. Because this problem has spherical symmetry, there are
``zero modes''.  One can rotate the charge from the north pole and
leave it in a displaced static position. A static rotation of
this type is equivalent in the soliton case to a non-exotic
``excitation'' (although in this toy model all non-exotic states
are degenerate with the ground state so ``excitation'' is a bit
of a misnomer).

Classical behavior associated with the exotic degrees of freedom
necessarily involves motion. Were there no velocity dependent
forces present due to the monopole, the existence of flat
directions would imply the existence of dynamical rotational modes
with the system slowly rotating around the entire sphere.
However, the magnetic monopole fundamentally alters this.  As soon
as the particle starts moving, the magnetic force acts to bend the
particle into a curved orbit and this curvature can be on a scale
much smaller than the radius of our sphere $R \sim N_c^0$; indeed
one can see self-consistently that the characteristic size of
such an orbit for states with small ``exoticness''(where $e = L
-|q g|$) will scale as $N_c^{-1/2}$ and, hence, does not go fully
around the sphere. If the orbit is localized in a region of this
size then at large $N_c$ it effectively stays in a region small
compared to $R$, the curvature of the sphere (which is order
$N_c^0$).  In this case the classical motion is effectively that
of a charged particle in a magnetic field, $\vec{B} = \hat{z}
g/R^2$, moving on a plane. Consider an orbit centered around the
north pole (one can always rotate your coordinate labels to do
this).  The position of the particle is then given approximately
by
\begin{equation}
x  \approx  r \cos (\omega t + \delta) \; \; \; y  \approx  r
\sin(\omega t + \delta) \; \; \; \omega \approx \frac{q g}{m R}
\label{classsol}\end{equation} where $r$ and $\delta$ are fixed
from the initial conditions and the corrections are of order
$1/N_c$.

Now suppose one wishes to quantize this classical circular
motion.  This is the familiar problem of Landau levels.  The
energy spectrum of such a system is of precisely the form of a
harmonic oscillator\cite{Landau}:
\begin{equation}
E =  (n + 1/2) \frac{| q B |}{m }  = (n + 1/2) \frac{|q g|}{m R^2}
\label{Landau}
\end{equation}
Equation (\ref{Landau}) can be derived using the Bohr-Sommerfeld
formula.  In doing this one finds that the radius of the orbits
are quantized to have
\begin{equation} r^2 \approx \frac{n+1/2}{|q B|} \approx
\frac{(n+1/2)R^2}{|q g|} \; ; \label{rtoy}
\end{equation}
the scaling rules in eq.~(\ref{toyscale}) then imply that the
radius scales as  $r \sim N_c^{-1/2}$.   This in turn, justifies
treating the problem as motion in the plane. Let us now look at
the excitation energies predicted from this semi-classical
quantization:
\begin{equation}
E_n -E_0 = n \frac{|q g|}{m R^2}\label{toysc} \; .
\end{equation}
The significant point is that eq~(\ref{toysc}) gives the same
excitation spectrum as in eq.~(\ref{toyspeca}) provided one
identifies the ``exoticness'', $e$, with the index $n$.

Of course, this toy model corresponds to rigid-rotor quantization
since there is a fixed moment of inertia.  It nevertheless
teaches us several things of importance which can be generalized
to situations where the moment of inertia is dynamical. i)  The
classical motion associated with the exotic excitation is not a
zero mode; no matter how slow the velocity the period of the orbit
is of order $N_c^0$. ii) The classical motion is bounded; for
energies corresponding  to low exoticness ($e \sim N_c^0$) the
typical size of the collective orbit is $N_c^{-1/2}$.  iii) In the
large $N_c$ limit the excitation spectrum can be understood by
purely semi-classical means.  Points i) and ii) indicate that as
far as the scales are concerned this problem ``looks like''
vibrational motion.

One aspect of point i) ought to be stressed.  The fact that this
dynamical mode is not a zero mode is essential. An apparently odd
feature is that the problem has two flat directions and one might
naively think  there ought to be two zero modes.  Where then does
the nonzero mode come from given the fact that there are only two
degrees of freedom in the problem? Actually, the issue is largely
semantic.  Consider a typical harmonic vibrational mode in one
dimension associated with the equation of motion $\ddot{x}= -
\omega_0^2 x $. The motion is given by $x=\alpha \cos(\omega_0 t)
+ \beta \sin(\omega_0 t)$ and is parameterized by two numbers
which are fit by two initial conditions, $\alpha =x(0)$ and
$\beta =\dot{x}(0)/\omega_0$.

 One typically describes this as one mode of
oscillation despite the need for two parameters to specify the
motion since one can always rewrite it as $x= A \cos (\omega_0 t
+ \delta)$, with $A=\sqrt{\alpha^2+\beta^2}$ and $\delta =
\tan^{-1}(\beta/\alpha)$.  Thus, the path followed depends only on
one parameter $A$, and the second parameter merely serves to
induce a phase shift in the single mode of motion. One might make
the following, somewhat pedantic, description: the motion consists
of two distinct modes; the general motion is then the
superposition of these two distinct modes. From this perspective
one can write the motion as two coupled first-order differential
equations for $x$ and $\dot{x}$ and define modes as solutions
where $x$ and $\dot{x}$ each evolve as $e^{ i \omega t}$ with the
same omega.  By defining modes this way it can be seen that there
are two modes in the harmonic problem given above---one with
$\omega=\omega_0$ and the other with $\omega=-\omega_0$---which
form a pair. For the problem of a single harmonic oscillator,
calling this a pair of modes with equal and opposite frequencies
may seem a bit artificial. However, in the toy problem of a
particle on a spherical shell moving in the field of a magnetic
monopole, this distinction is important. There are two degrees of
freedom and thus, in the sense given above, one expects four modes
to come in two pairs of equal and opposite $\omega$. This is
precisely what happens.  One pair of modes corresponds to fully
static configurations (one moves the charge in either the $x$ or
$y$ direction to a new position and leaves it there at rest) and
these  correspond to two zero modes which we view as a single pair
of modes.  The other pair of modes correspond to the charge
orbiting either clockwise or counterclockwise corresponding to
$\omega=\pm \frac{|q g|}{m R^2}$. However, while we still have
two pairs of modes the existence of a velocity dependent force
means that the pairs do not correspond to a mode associated with
initial displacement in a direction paired with an initial
velocity in the same direction. This means that although this
problem has two zero modes associated with displacements, it does
not have two pairs of zero modes but only one.

\subsection{Coupled particles in a monopole field\label{coupled}}

While the problem in the previous section gives some insights
into the scales of the problem and the nature of modes and
semi-classical quantization, it cannot answer the fundamental
question about whether rigid-rotor quantization is generically
valid since by construction the problem is a rigid rotor.
Accordingly we need a model where the underlying dynamics is not
rigid; {\it i.e.,} a model where the analog of the soliton has
some nontrivial internal dynamics.  Moreover, because the central
concern is mixing between collective and internal modes, it is
important that the internal degrees of freedom have excitations
which have the same quantum number numbers as the exotic motion.
The simplest model of this sort is a generalization of the model
studied in the previous section.  Consider a non-relativistic
theory of two particles confined to the surface of a sphere.  For
simplicity we will take them to have equal mass, $m$.  At the
center of the sphere is a magnetic monopole of strength $g$.  The
two particles have different charges---which  we will take to be
$q$ and zero---so they interact differently with the monopole.
Finally, the two particles interact with each other via a
quadratic potential $V_{\rm int} = k |\vec{r_1} -\vec{r_2}|^2/2$,
where $k$ acts as a spring constant.  To create an analogy with
SU(3) solitons at large $N_c$ the following scaling rules must be
imposed  for the parameters:
\begin{equation}
q \sim N_c^0 \; \; \; \; R \sim N_c^0 \; \; \; \; g \sim N_c^1 \;
\; \; \; m \sim N_c^1 \; \; \; k \sim N_c^1 \; .\label{toyscale2}
\end{equation}

The analog of the classical soliton is a static configuration
which solves the classical equations  of motion.  The solution is
simple: the particles are on top of each other and we can take
their position to be at the north pole.  The rigid-rotor
quantization of this system is quite trivial. The classical
``soliton'' is constrained to move coherently with the two
particles in their classical ground state ({\it i.e.,} with the
two particles on top of each other).  This reduces immediately to
the classical motion of a single particle of mass $2m$ and charge
$q$ and one can immediately read off the excitation by making the
appropriate substitutions in eq.~(\ref{toyspeca}):
\begin{equation}
 E_e^{\rm rigid} - E_0 = e \omega_r \; \; \; {\rm with} \; \;  \; \omega_r \equiv \frac{ |q
g|}{ 2 m R^2}   \label{toyrigid}  \; ,
\end{equation}
where as before $e$ is a non-negative integer.

On the other hand, we can follow the correct full procedure of
finding classical time-dependent solutions and then quantize this
motion semi-classically.  All small amplitude motion can easily
be found via a linearization of the equation of motion; this is
sufficient provided the quantization keeps the motion within the
regime of validity of linear response.  As before, this can be
checked {\it a posteriori}.  If one is in the small amplitude
regime the problem again reduces to motion on a plane in the
presence of a magnetic field.  Thus we can parameterize the two
degrees of freedom for each particle by its $x$ and $y$
coordinates. Following the discussion in the previous subsection,
it is useful to write first-order equations of motion for the
particles and their associated velocities:
\begin{widetext}
\begin{equation}
\frac{d}{d t} \left( \begin{array}{c} x_1\\ y_1\\ x_2 \\
y_2\\\dot{x}_1\\\dot{y}_1\\\dot{x}_2\\\dot{y}_2 \end{array}
\right ) \, = \, \left ( \begin{array}{c c c c c c c c}
 0 & 0 & 0 &  0 & 1 & 0 & 0 & 0 \\
 0 & 0 & 0 &  0 & 0 & 1 & 0 & 0 \\
 0 & 0 & 0 &  0 & 0 & 0 & 1 & 0 \\
 0 & 0 & 0 &  0 & 0 & 0 & 0 & 1 \\
 \frac{-\omega_v^2}{2} & 0 & \frac{\omega_v^2}{2}& 0 & 0 & 2 \omega_r & 0 & 0 \\
 0 & \frac{-\omega_v^2}{2} & 0 & \frac{\omega_v^2}{2} & 2 \omega_r & 0 & 0 & 0  \\
\frac{\omega_v^2}{2} & 0 & \frac{-\omega_v^2}{2}& 0 & 0 & 0 & 0 & 0 \\
 0 & \frac{\omega_v^2}{2} & 0 & \frac{-\omega_v^2}{2} & 0 & 0 & 0 & 0  \end{array} \right )
 \left( \begin{array}{c} x_1\\ y_1\\ x_2 \\
y_2\\\dot{x}_1\\\dot{y}_1\\\dot{x}_2\\\dot{y}_2 \end{array}
\right )\; \; \; {\rm with} \; \; \; \omega_v=\sqrt{\frac{2
k}{m}} \; \; \; {\rm and} \; \; \; \omega_r=\frac{g q}{2 m R^2} \;
\; . \label{m}\end{equation}
\end{widetext}
The quantities $\omega_v$ and $\omega_r$ have simple physical
interpretations: $\omega_v$ is the vibrational frequency of the
two particles when the monopole field is turned off and
$\omega_r$ is the Landau frequency assuming that rigid-rotor
quantization is valid. From the scaling rules in
eq.~(\ref{toyscale2}), ones sees that $\omega_r \sim N_c^0$ and
$\omega_v \sim N_c^0$. The normal mode frequencies are $-i$ times
eigenvalues of the matrix on the right-hand side of eq.~(\ref{m}).

\begin{figure}
\includegraphics{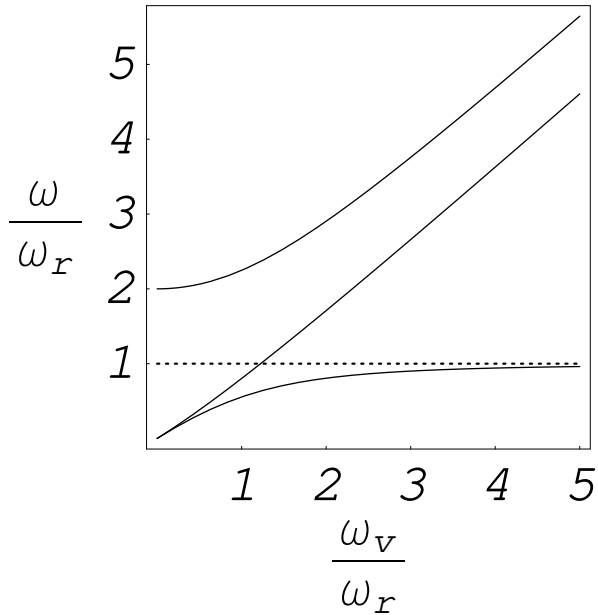}
\caption{Non-vanishing normal mode frequencies for the model in
subsection \ref{coupled} in units of of $\omega_r$ as a function
of the ratio $\frac{\omega_v}{\omega_r}$ with $\omega_r$ and
$\omega_v$ defined in eq.~(\ref{m}).  The strongly coupled limit
corresponds to $\frac{\omega_v}{\omega_r} \rightarrow \infty$. The
dashed line represents the prediction for the lowest excitation
energy in rigid-rotor quantization. This prediction agrees with
the full calculation only in the strong coupling regime. }
\label{omega}
\end{figure}

The matrix can be diagonalized explicitly and it is found that, as
expected, the eight modes group into four sets of pairs of modes
with equal and opposite frequencies.  There is one pair of zero
modes which corresponds to static configurations in which both
particles are displaced by the same amount. Analytic expressions
can be obtained for the frequencies of the three pairs of
non-zero modes. However, these expressions are quite cumbersome.
Since the modes depend only on $\omega_v$ and $\omega_r$, it is
useful to express the frequencies as  multiples of $\omega_r$ and
to express results as a function of $\frac{\omega_v}{\omega_r}$.
A plot of the three (positive) frequencies for these pairs is
given in fig.~\ref{omega}. These modes can be semi-classically
quantized and the lowest-lying excitation associated with the
motion is  $\omega$ above the ground state.  As before, the
quantized orbits correspond to velocities of order $N_c^{-1/2}$
and displacements of order  $N_c^{-1/2}$ which self-consistently
justifies the neglect of the curvature at large $N_c$.

The results in fig.~(\ref{omega}) are quite striking.  One sees
that the rigid-rotor result is reproduced for one of the modes
only in the limit where $\frac{\omega_v}{\omega_r} \rightarrow
\infty$.  This can be verified explicitly: expanding the analytic
expression for the appropriate eigenvalue which we denote as
$\omega_c$ as a series in $\frac{\omega_r}{\omega_v}$ yields
\begin{equation}
\omega_c = \omega_r \left( 1 - \left(\frac{\omega_r}{\omega_v}
\right )^2 + \left(\frac{\omega_r}{\omega_v} \right )^4 - 3
\left(\frac{\omega_r}{\omega_v} \right )^8 \, +  \, \cdots  \right
) \; \; . \label{modeex}
\end{equation}
Clearly this series converges onto $\omega_r$ as
$\frac{\omega_v}{\omega_r} \rightarrow \infty$ and deviates from
this asymptotic value at finite values of the ratio. This result
is very easy to understand physically. The limit
$\frac{\omega_v}{\omega_r} \rightarrow \infty$ corresponds to
tight binding; the ratio diverges as $k $, the strength of the
inter-particle interaction goes to $ \infty$ with all other
parameters held fixed. Of course, in the tight binding limit the
system will act like a single coherent particle of mass $2 m$ and
charge $q$ and one recovers the rigid-rotor result.

Moving away from the tight binding limit the particles no longer
move coherently.  Suppose one were to provide an initial
condition in which both the particles were given an equal kick so
that at $t=0$ they both had the same initial velocity and had no
initial separation.  As they move, they feel different magnetic
forces due to the differing charges and hence begin to move
apart. At this point the Hook's law potential between the particle
adds a new restoring force and the ``collective'' and vibrational
motion now mix and the modes differ from the rigid-rotor modes.
Indeed, as one moves to the opposite limit the result is also
easy to understand analytically. As $k \rightarrow 0$ with all
other parameters held fixed the two particles only weakly
interact with each other; the charged particle is expected to make
Landau oscillations essentially unencumbered by the other
particle. Since the single-particle mass is exactly half the mass
of the coherent system in the tight binding limit the frequency
of the Landau oscillation is exactly double that case.  As can be
seen from fig.~\ref{omega}, this is precisely what happens: one
of the modes goes to 2 $\omega_r$ in this extreme weak binding
limit. The other frequencies become small due to the lack of a
significant restoring force.

The important thing to note is that scaling rules in
eq.~(\ref{toyscale2}) imply the ratio $\frac{\omega_v}{\omega_r} $
is generically of order $N_c^0$. Thus, the large $N_c$ does not
automatically force one into the tight binding limit.  This in
turn means that large $N_c$ by itself does not push one into the
regime of validity of the rigid-rotor quantization.  We see
explicitly that rigid-rotor quantization is not automatically
justified at large $N_c$.

It is reassuring that the analysis of this simple toy model
reproduces the general arguments about time scales discussed in
subsection \ref{tscales}.  The rigid-rotor quantization is seen
to be  justified only when $\omega_v \gg \omega_r$. This  is the
situation when there is a scale separation between the collective
and vibrational motion and Born-Oppenheimer reasoning applies.
However, as just noted, large $N_c$ QCD does not imply that the
dynamics is in this regime.

This model is also useful in clarifying some key issues. It is
well known in typical soliton models that zero modes associated
with broken symmetries are orthogonal to vibrational degrees of
freedom at small amplitude and do not mix\cite{sol}.  It was
precisely this fact that was used to justify rigid-rotor
quantization. However, we see that for this toy problem the
rigid-rotor quantization fails.  The reason for this failure is
easy to isolate.  Although there are two flat directions, the
presence of velocity-dependent forces induced by the monopole
implies that there is only one pair of zero modes and not two.
This second ``would be'' zero mode scales with $N_c$, an ordinary
vibrational mode, and acts as an ordinary vibrational mode.
Nothing prevents it from mixing with other vibrational modes and
indeed it does. This can be seen explicitly by looking at one of
the eigenvectors of the ``would be'' zero mode.  Again, the
expression  is long and cumbersome in general, but can be written
in a Taylor series in $\frac{\omega_r}{\omega_v}$.  To second
order, the motion associated with this normal mode is given by
\begin{widetext}
\begin{equation}
 \left( \begin{array}{c} x_1\\ y_1\\ x_2 \\
y_2\\\dot{x}_1\\\dot{y}_1\\\dot{x}_2\\\dot{y}_2 \end{array}
\right ) = \Re{\rm e} \left ( A \exp(i \omega_c t) \left \{
\left( \begin{array}{c} 1\\ i\\ 1 \\
i \\0\\0\\0\\0 \end{array}
\right ) + \left( \frac{\omega_r}{\omega_v} \right ) \left( \begin{array}{c} 0\\ 0\\ 0 \\
0\\ i\\ -1\\ -i\\ -1\end{array}
\right )  + \left ( \frac{\omega_r}{\omega_v}\right )^2 \left( \begin{array}{c} -1\\ i\\ 1 \\
-i\\0\\0\\0\\0 \end{array} \right ) \, + \, \cdots \right \}
\right ) \label{modemix}\end{equation}
\end{widetext}
where $\Re{\rm e} $ indicates the real part, $A$ is a complex
constant fixed by initial conditions and $\omega_c$ is the normal
mode frequency given approximately in eq.~(\ref{modeex}).  Note
that although the leading term in the expansion has particles one
and two moving together in a coherent manner (making a circular
orbit), the correction terms do not. Again, we should recall that
the large $N_c$ limit does not drive one to the limit where
$\frac{\omega_r}{\omega_v}$ goes to zero so that large $N_c$ does
not force these correction terms to be small: at leading order in
a $1/N_c$ expansion, the collective motion has mixed with
vibrational motion.

Returning to the general considerations at the beginning of this
section, we see that for models such as this one which contain
first order time derivatives, the classical intrinsic vibrational
motion can mix with the collective rotational motion.  This in
turn implies that the collective motion can not be isolated and
quantized separately.

\subsection{SU(3) soliton models}

The preceding subsection shows explicitly why rigid-rotor
quantization fails for that toy model. But we are interested in
SU(3) soliton models. The issue is whether the same type of
behavior is seen in SU(3) models. The important point is that the
behavior seen in the toy model should be generic. Any model with
``exotic'' motion of order $N_c^0$ and vibrational motion of the
same order can be expected to mix in the absence of some symmetry
keeping them distinct.

In fact, we know from the analysis of IKOR that this same behavior
is seen for SU(3) solitons\cite{IKOR}.  They study the classical
motion around the soliton at quadratic order.  At large $N_c$
 this quadratic order is sufficient since all nonzero kaon modes when
quantized have $K/f_\pi \sim N_c^{-1/2}$ corresponding to
localized and nearly harmonic motion. While for the non-exotic
states at large $N_c$ and exact SU(3) symmetry, they find a
rotational zero mode as seen in collective quantization, they
also note that there is no exotic mode with a frequency equal to
the excitation given in rigid-rotor quantization.  However, if
the classical motion really had separated into collective and
intrinsic motion as assumed in rigid-rotor quantization there
would have been a classical mode at the rigid-rotor frequency.
Thus the SU(3) solitons do behave in the same way as the toy
model.

\section{The critique of the time scale argument}

As noted in the introduction, the conclusion that the rigid-rotor
quantization for exotic states is not justified by large $N_c$
QCD is not universally accepted.  Recently Diakonov and Petrov
(DP) \cite{DP}criticized the argument given in ref.~\cite{Coh03}
(and presented here in subsection \ref{tscales} of this
paper)---{\it i.e.,} the argument based on time scales; using the
logic of this critique they concluded that rigid-rotor
quantization was valid for ``exoticness'' of order $N_c^0$ and
only breaks down when the ``exoticness'' is large enough to
substantially alter the moment of inertia (which occurs at order
$N_c^1$)\cite{DP}. In this section we will discuss this critique
and argue that it is erroneous.

The critique has three parts: i) An argument that relevant time
scale for exotic excitations is of order $N_c^{1}$ and not order
$N_c^0$; ii) a general treatment of rotational-vibrational mixing
in  SU(3) solitons in which it is asserted that the mixing is
small at large $N_c$; and iii) a toy model to illustrate the
argument.

Let us begin with the discussion of point i).  DP argue correctly
that in the  rigid-rotor quantization the characteristic angular
velocity is given by
\begin{equation}
v_{char} \sim  \frac{\sqrt{\sum_{A=4}^7 {\hat{J}_A'}{}^2}}{ I_2}
\sim {\cal O}(\sqrt{\frac{e}{N_c}}) \label{vchar}
\end{equation}
where $e$ is the ``exoticness'' .  This is slow at large $N_c$
when $e \sim N_c^0$. From the behavior in eq.~(\ref{vchar}), it is
asserted that the characteristic time scale is of order
$N_c^{-1/2}$\cite{DP}, in contradiction to the claims in
ref.~\cite{Coh03} and subsection \ref{tscales} that the time
scale is of order $N_c^0$. This assertion is, on its face,
paradoxical. If true, general semi-classical considerations as
seen in Bohr's correspondence principle would imply that the
quantal excitation energies associated with this be of order
$1/N_c$ and not $N_c^0$, as, in fact, they are .  DP attempt to
resolve this paradox by stating that the rotation is not
semi-classical because of the quantization condition in
eq.~\ref{quantcond} which arises from the Wess-Zumino term; since
the Wess-Zumino term is a full derivative term it should be
thrown out classically and is thus not suitable to semi-classical
arguments. However, this resolution does not hold up. As seen
explicitly in subsection \ref{mono} the problem of ``exotic''
motion of a charged particle in a monopole field can be quantized
explicitly by quantizing the Landau orbits and the semi-classical
result agrees with the exact answer at the analog of large $N_c$.

Given the problem with the resolution of this paradox by DP, how
is one to reconcile eq.~(\ref{vchar}) with a characteristic time
of order $N_c^0$?  The answer is trivial and was discussed in
detail in subsection \ref{tscales}: the velocity can be of order
$N_c^{-1/2}$ and the time scale $N_c^0$ provided the motion is
localized to a region of order $N_c^{1/2}$.  This is precisely
what happens and this is verified in the model in subsection
\ref{mono}.  In summary, part i) of this critique appears to be
without foundation.

Next consider part ii).  The analysis here closely parallels the
original derivation of rigid-rotor quantization.  Let us
recapitulate the salient points.

The analysis is based on a soliton which corresponds to the local
minimum of an effective action $S_{\rm eff}[\pi(x)]$ (where $\pi$
represents a dimensionless pion field---namely, the usual pion
field divided by $f_\pi$); the action is proportional to $N_c$.
The classical configuration $\pi_{\rm class}(x)$ which minimizes
the effective action gives the soliton profile and the  moments
of inertia $I_{1,2}$ are computed at this minimum. One finds that
the classical soliton mass, ${\cal M}_0$, and the moments of
inertia, $I_{1,2}$, are all proportional to $N_c$.

The effective action  may be expanded about the classical minimum
at second order in the fields; it is given by:
 \begin{equation} {\cal E}_{\rm eff}[\pi_{\rm class}+\delta\pi] =
{\cal M}_0+\frac{1}{2}\delta\pi\,W[\pi_{\rm
class}]\,\delta\pi+\ldots \label{S2}
\end{equation}
where $W$ is an operator for any given external field $\pi_{\rm
class}$. Since the dimensionless fields scale as $N_c^0$, $W$ is
of the same order  $S_{\rm eff}$, namely, $N_c^1$.  This in turn
implies that the harmonic fluctuations scale as $\delta\pi(x)
=O(1/\sqrt{N_c})$. The spectrum  of $W$ and its eigenmodes both
scale as $N_c^0$.  Clearly $W$ has zero modes which are related to
symmetry breaking in the classical solution. For these models
this includes both translations and rotations.

Up to this point the analysis seems reasonable, although one
might quibble that the object on the left-hand side of
eq.~(\ref{S2}) should be considered an effective Lagrangian
rather than an energy function since no Legendre transformation
has yet been made.  In any event, the next steps \cite{DP} are
based on a set of  assumptions which are quite problematic:
 `` {\it The quantization
of rotations (which are large fluctuations as they occur in flat
zero-mode directions) leads to the rotational spectrum discussed
in the previous section {\rm [that is rigid-rotor quantization]}.
The vibrational modes are orthogonal to those zero modes.}''
Given theses assumptions,  it is shown that vibration-rotational
coupling only becomes important when the moment of inertia
changes substantially which occurs at $e \sim N_c^1$.

However, the assumption that the exotic motion is associated with
zero modes and thus do not mix with vibrational motion due to
orthogonality was assumed without proof to be true in analogy
with problems without a Wess-Zumino term.  This assumption doe not
appear to be correct. In the first place, the classical motion
associated with exotic motion is not associated with a zero mode
due to the presence of the Wess-Zumino term. This was shown in
subsection \ref{tscales} where it was seen that classical zero
modes are associated with quantum states that have level spacing
which go to zero as the large $N_c$ (classical) limit is taken.
Exotic states are not in this class.  Moreover, in subsection
\ref{mono} it was shown explicitly that the classical motion
associated with exotic states were not zero modes; rather, they
were Landau orbits with a nonzero frequency.  Secondly, for
exotic quantum numbers, the collective motion does mix with the
vibrational motion at leading order as is shown explicitly in
eq.~(\ref{modemix}). Thus, the assumption underlying this general
part of the critique seems to be wrong.

Finally, we consider part iii) of the critique.  This consists of
studies with a toy model with both vibrational and rotational
degrees of freedom and with an analog of a Wess-Zumino term. The
model considered was a charged particle moving in the field of a
magnetic monopole and subject to a spherically symmetric potential
with a minimum at $r=R$.  In this model, it is shown that
rigid-rotor quantization only becomes inaccurate when the moment
of inertia is altered substantially---an effect which occurs when
the exoticness is of order $N_c$. However, this model is a very
poor analog of the problem of interest.  Recall that the danger
posed to rigid-rotor quantization is the mixing of the collective
modes with vibrational modes {\it which carry the same quantum
number}.  Note that in the toy model considered here the exotic
motion carries angular momentum (also note that the exotic states
correspond to states with different J) while the only vibrations
allowed in this model are radial vibrations which do not carry
angular momentum.  In this model there is nothing for the
collective modes to couple to at lowest order and rigid-rotor
quantization works.  However, as soon as the model is rich enough
to include vibrational degrees of freedom with the same quantum
numbers as the collective motion (as, for example, in the model of
subsection \ref{coupled}), the rigid-rotor quantization fails.

To summarize this section, DP make a three part critique at the
analysis of time scales given in subsection \ref{tscales} and in
ref.~\cite{Coh03}.  However, the first two parts of the critique
are based on faulty assumptions, while the third part is based on
a model which is not an analog of the relevant problem.

\section{Discussion}

This paper presents strong evidence that rigid-rotor quantization
is not justified on  the basis of large $N_c$ considerations. The
important issue is what this tells us about the nature of the
$\theta^+$ and other possible exotic states.

One possibility is that the analysis based on the rigid-rotor
quantization is, in fact, well justified despite the arguments
given in this paper.  Note that this paper {\it does not} show
that rigid-rotor quantization is necessarily invalid but rather
that it is not justified due to large $N_c$ QCD.  It remains
possible that it is justified due to some other reason.  (For
example, in the toy model in  subsection \ref{coupled} rigid-rotor
quantization was justified if the parameters of the model had
$\omega_v \gg \omega_r$).  This would be most satisfactory in
that the very successful prediction of the phenomenology based on
rigid-rotor quantization would remain.  However,  if correct, it
raises a very important theoretical question: namely, what
justifies the rigid-rotor quantization?  A second possibility is
that the rigid-rotor quantization is not justified.  In this
case, the accurate prediction of the mass and the prediction of a
narrow width in rigid-rotor quantization would have to simply be
dismissed as fortuitous.

In either case, large $N_c$ QCD by itself does not appear to allow
one to understand the structure of this state.  In this respect,
the $\theta^+$ is quite unlike the $\Delta$ and is like the more
typical excited baryons.  As with such baryons it may well be that
the best phenomenological treatments may be based on models whose
connections to QCD are quite tenuous.

The lack of validity of the rigid-rotor quantization does have one
important phenomenological consequence.  If the exotic states
were of a collective character one could not justify treating
meson-baryon scattering in exotic channels via a simple linear
response theory in the context of chiral soliton models. After
all, one cannot use linear response to describe pion-nucleon
scattering at the $\Delta$ resonance in the Skyrme
model\cite{KarMat,MatPes}.  However, because these exotic
resonances are of a vibrational character, linear response is
justified. This is useful in and of itself to describe
scattering. Moreover, by imposing the $I=J$ rule on such
scattering one can predict that the $\theta^+$ has low-lying
partners (at least) at large $N_c$\cite{CohLebTheta}.  These
partners are related in much the same way as the $\Delta$ is to
the nucleons except that both states have widths which are of
order $N_c^0$. The quantum numbers of such partner states are
enumerated in ref.~\cite{CohLebTheta}.

The author acknowledges Rich Lebed and Pavel Poblytisa for
insightful suggestions about this work. The support of the U.S.
Department of Energy for this research under grant
DE-FG02-93ER-40762 is gratefully acknowledged. \\

\appendix
\section{Quark Based Derivation of the Wess-Zumino Constriant \label{const}}
 In Skyrme-type models eq.~(\ref{quantcond}), which constrains the
 allowable representations follows directly from the Wess-Zumino term
 (which topology fixes to have a strength which is integer and can be identfied
identified with $N_c$ \cite{Wit2}). At a more pedestrian level,
it can also be easily understood at the quark level for models
with explicit quark degrees of freedom. In the body-fixed frame
the baryon number of the unrotated hedgehog is associated with
the SU(2) sub-manifold. The body-fixed hypercharge is also
associated with this sub-manifold.  One can relate the body-fixed
hypercharge to the body-fixed SU(3) generator as usual so that
$Y'= -2 {J'}_8/ \sqrt{3}$. The baryon number, hypercharge and
strangeness are related linearly.  The appropriate relation for
arbitrary $N_c$ is
\begin{equation}
Y = \frac{N_c B}{3} + S \label{hyper} .
\end{equation}
Note that eq.~(\ref{hyper}) does not coincide with the familiar
relation $Y=B+S$ except for $N_c=3$. For arbitrary $N_c$
eq.~(\ref{hyper}) may be obtained from the known hypercharges of
up, down and strange quarks:
\begin{equation}
 Y_u = 1/3  \; \;Y_d =1/3 \; \; Y_s=-2/3\; .
\end{equation}
 (These are the standard hypercharge assignments for quarks
at $N_c=3$.  It is straightforward to see that these assignments
must hold for any $N_c$ provided hypercharge is isosinglet and
traceless in SU(3) and has the property that the hypercharge of
mesons is equal to the strangeness.) Each quark  carries a baryon
number of $1/N_c$ while the strangeness is zero for u and d
quarks and -1 for s quarks.  The combination of hypercharge,
strangeness and baryon number assignments can only be satisfied if
eq.~(\ref{hyper}) holds. Finally, observe that in the  body-fixed
frame, the SU(2) sub-manifold by construction has zero
strangeness; thus, eq.~(\ref{hyper}) requires that $Y' = N_c B/3$
and the quantization condition in eq.~(\ref{quantcond})
immediately follows.

\end{document}